\documentclass{article}
\usepackage{spconf,amsmath,graphicx}
\usepackage{url}
\usepackage{enumitem}
\setlist{nosep, leftmargin=14pt}

\usepackage{mwe} % to get dummy images

\title{Spatially-Preserving Flattening for Location-aware\\ Classification of Findings in Chest X-rays}
\name{Neha Srivathsa$^{1}$, Razi Mahmood$^{2}$, Tanveer Syeda-Mahmood$^{3}$}
\address{$^{1}$Stanford University, $^{2}$University of California, Berkeley, $^{3}$IBM Almaden Research Center}
\begin{document}
%\ninept
%
\maketitle
\begin{abstract}
Chest X-rays have become the focus of vigorous deep learning research in recent years due to the availability of large labeled datasets. While classification of anomalous findings is now possible, ensuring that they are correctly localized still
remains challenging, as this requires recognition of anomalies within anatomical regions. Existing deep learning networks for fine-grained anomaly classification learn location-specific findings using architectures where the location and spatial contiguity information is lost during the flattening step before classification. In this paper,  we present a new spatially preserving deep learning network that preserves location and shape information through auto-encoding of feature maps during flattening. The feature maps, auto-encoder and classifier are then trained in an end-to-end fashion to enable location aware classification of findings in chest X-rays. Results are shown on a large multi-hospital chest X-ray dataset indicating a significant improvement in the quality of finding classification over state-of-the-art methods.
\end{abstract}
\section{Introduction}
\label{sec:intro}
Chest X-rays are the most common imaging exams done in emergency rooms and intensive care units in hospitals. With the availability of large open source datasets labeled for selected anomalies\cite{Johnson2019,WangXPengYLuLLuZBagheriM2017}, they have become the focus of vigorous deep learning model research in recent years\cite{IrvinJRajpurkarPKoM2019,WangXPengYLuLLuZBagheriM2017,syeda-mahmood2020, Wu2020}. The goal of this research is to produce automated radiology reports, which requires recognition of fine-grained findings that describe the laterality as well as location of findings (e.g. "left basal atelectasis"). While it is possible to do detection and classification with modern deep learning networks in computer vision\cite{Lin2017,Carion2020}, fine-grained finding recognition is different as it requires the anomaly to be located within the correct anatomical reference (e.g. pneumothorax in left apex). Further, certain anomalies only appear in specific anatomical regions, requiring the modeling of these constraints (e.g. opacities are seen in lungs). Similarly, capsule networks\cite{Sabour2017} that are meant to capture relative spatial relationships of parts of an object in multiple poses are not very applicable for chest X-rays since anomalies need to be localized within relevant anatomical areas that in turn, also need to be identified.

Large-scale fine-grained finding recognition in chest X-rays is a relatively recently posed problem. The current methods to handle such fine-grained findings use 3 main approaches: (a)  incorporating an anatomical atlas for localization prior to classification\cite{Agu2021}, (b) using soft attention with gradCAM within classifiers themselves\cite{Kashyap2020}, and (c) direct classification using detailed finding labels that already reflect the location and laterality in the label names\cite{Wu2020}. In a recent work, a fixed number of anatomical zones in a chest X-ray were localized through bounding boxes within atlas-guided anatomical segmentation\cite{Wu2020AutomaticBB}. The alignment of these bounding boxes for new chest X-rays required ad hoc registration.  Later work used the training bounding boxes per anatomical region as supervision data to build a faster RCNN network\cite{Ren2015} for detection. The inter-relationship features between regions was then extracted using a graph convolutional network and a multi-class classifier was applied to each region to classify anomaly labels within each region\cite{Agu2021}. This two-stage approach, while reasonable, can lead to potential incorrect combinations of anomalies with anatomical locations as all regions are examined for all anomalies. In the second class of approaches, soft attention using heat maps was generated through gradCAM operators in classifier networks\cite{Kashyap2020}, or obtained using occlusion sensitivity as a measure of localization\cite{Islam2017}. Others have adopted hard attention applied through regions of interest obtained by a prior region segmentation algorithm based on U-net\cite{Ronneberger2015} for lungs\cite{Wu2020AutomaticBB}. Soft attention approaches are based on the hope that the classifier is looking in the relevant region for the label which can only be verified in a post-hoc way through operators like GradCAM. The hard attention models, on the other hand, rely on the accuracy of deep learning-driven region segmentation, which is difficult even for large regions such as lungs, in the presence of severe anomalies. Both atlas-based and attention-based approaches have only been attempted on a limited class of findings. Finally, the third class of approaches uses a pure classifier but increases the granularity of the labels to cover location and laterality information within anomaly labels. Specifically, companion radiology reports were used to derive detailed fine-grained labels which were then directly learned using a custom deep learning model designed for multi-label classification\cite{Wu2020,tanveer2020}. As the model used whole image inputs and a flattening layer for classification that lost spatial contiguity information, the fine-grained classification performance was low (weighted average AUC of 0.73). Due to the bundled approach, an error in class label could point to gross errors in localization as well. Nevertheless, the classifier approach has been the only one so far to cover a large spectrum of findings suitable for the ultimate application of automatic report generation. 

In this paper, therefore, we adopt the classifier approach for fine-grained labels, but address the limitation of existing classifier deep learning networks that lose spatial contiguity during the flattening step prior to classification. Specifically,  we present a new spatially preserving deep learning network that preserves location and shape information through auto-encoding of feature maps during flattening. Results are shown on a large multi-hospital chest X-ray dataset indicating a significant improvement in the quality of finding classification over state-of-the-art methods without requiring detailed anatomy segmentation or large-scale region annotation.  
%In this paper, therefore, we adopt the classifier approach by capturing the location and laterality information in the labels. However, to increase the sensitivity and specificity of such an approach, we develop a new spatial location information preserving deep learning network for location-aware finding classification. Results are shown on a large multi-hospital chest X-ray dataset indicating a significant improvement in the quality of findings classified over state-of-the-art methods.  
\section{Preserving spatial information during classification}
\label{sec:format}
To understand the key idea behind our approach, consider the architecture of most deep learning networks for image classification, such as VGG-16\cite{vgg16}or Resnet101\cite{HeKZhangXRenS2009}. Typically, they have a feature extraction stage formed from layers of convolutional filtering and pooling operations followed by an intermix of fully connected and dropout layers for classification\cite{vgg16,HeKZhangXRenS2009}. Although the feature extraction operations using convolutional filters are applied at a pixel or super-pixel level, these are ultimately flattened into vectors before proceeding to the classification stage, resulting in the loss of spatial contiguity and layout information. %Consider the network of Figure 6 which shows a recent network proposed for fine-grained label classification in chest X-rays(13). Here complex feature extraction operators using feature pyramid network for multi-resolution image analysis and dilation operations were used to ultimately generate 128 feature maps of size 64 x 64 which was then flattened using global square pooling into a single 16384-vector causing spatial locality and order to be lost since values from all positions were averaged.
% with a suitable PostScript file name.
\begin{figure}[htb]
\begin{minipage}[b]{1.0\linewidth}
  \centering
  \centerline{\includegraphics[width=8.5cm]{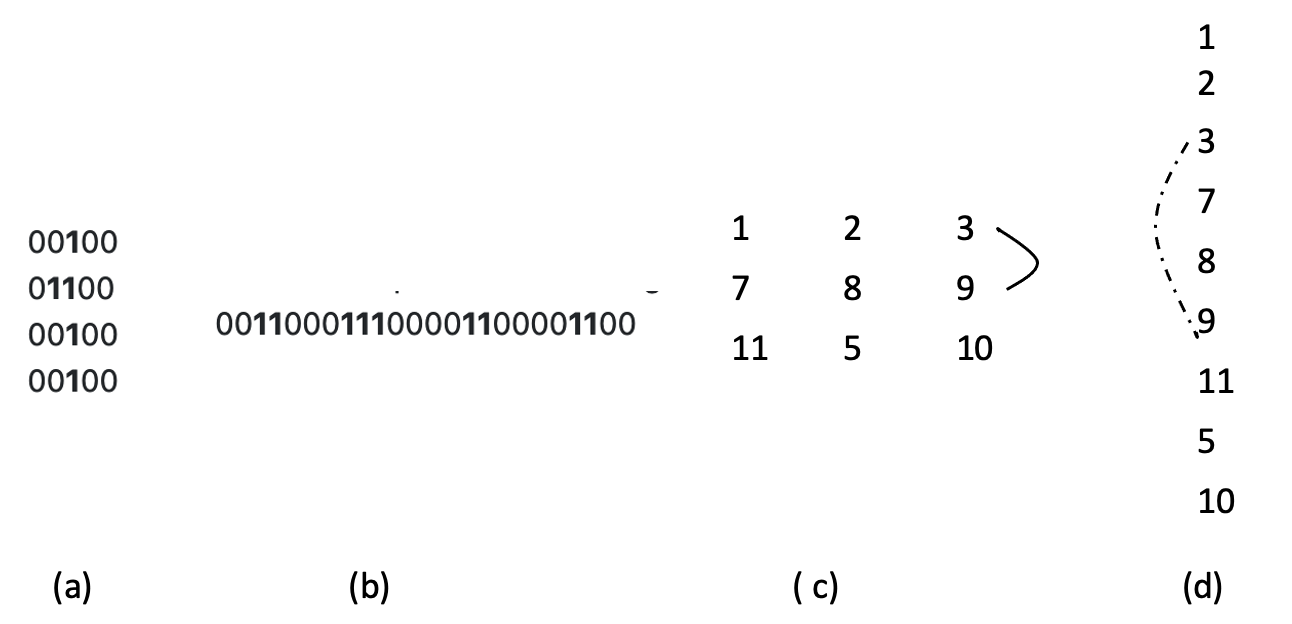}}
  \centerline{(a) Result 1}\medskip
\end{minipage}
\caption{Illustration of loss of spatial contiguity during flattening - simple examples.}
\label{simplefig}
\end{figure}
This can be easily seen through a simple example in Figure~\ref{simplefig}, where the 2D layout makes it trivial to see the arrangement as representing the number 1 in Figure~\ref{simplefig}a, in comparison to its flattened representation in Figure~\ref{simplefig}b. Similarly, in Figure~\ref{simplefig}c and d, the spatial adjacencies of the intensity values across a column are lost in the flattening, resulting in a loss of spatial information.
{\em If the spatial layout information could be preserved during the flattening step, it would be possible to distinguish between location-specific anomalies better during classification}. This is the key idea proposed in this paper. Specifically, we preserve spatial information in two ways, (i) by capturing the shape information conveyed in the filtered representations, and (ii) by keeping the identities of the filters separate during the flattening. We now describe this in detail. 
\subsection{Spatially-preserving flattening}
\begin{figure}[htb]
\begin{minipage}[b]{1.0\linewidth}
  \centering
  \centerline{\includegraphics[width=8.5cm]{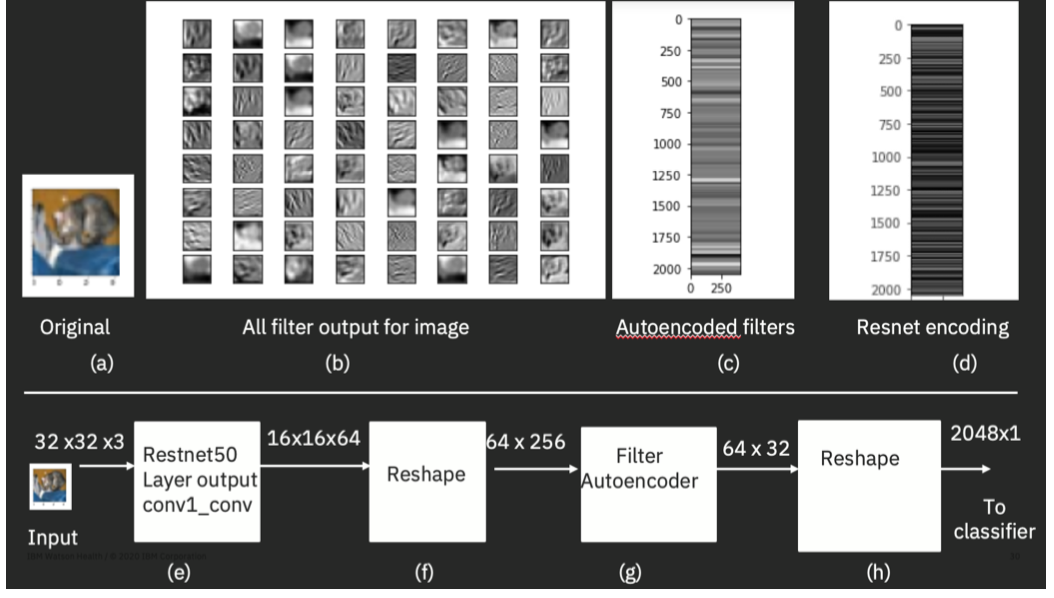}}
\end{minipage}
\caption{Illustration of spatially preserving flattening in ResNet50v2 using the filters from the second layer.}
\label{resnet}
\end{figure}

 The convolutional filters in a deep learning network learn to detect image characteristics at specific locations in either the direct image (at the input layer) or its successive abstractions obtained through pooling operations capturing more global characteristics. The result of applying these filters to an image generates an activation map, also known as a feature map, highlighting the relevant features detected or preserved in the input by the filter. While feature maps close to the input detect small or fine-grained detail, and feature maps close to the output of the model capture more general features, each feature map can be seen as defining a shape.  Figure~\ref{resnet} illustrates this for a Resnet50\cite{HeKZhangXRenS2009} architecture. Here the feature maps produced by applying the filters in Layer 2 ( $conv1conv$ layer) consist of 64 filters of 16x16 each. The 64 feature maps produced using an instance of Resnet50 pretrained on Imagenet from the input image of Figure~\ref{resnet}a are shown in Figure~\ref{resnet}b.
 {\em Since auto-encoders are known to form a low-dimensional representation or encoding of shapes, we can represent each such feature map using an auto-encoder and inherently preserve its spatial information}. Further, by concatenating all the encodings of the feature maps, we form a new flattening vector that separately retains the shape information from the individual feature maps. A fixed ordering can be used among the feature maps to define the layout of their encodings. Figure~\ref{resnet}c shows the resulting encoding vector generated from the auto-encoding of feature maps of Figure ~\ref{resnet}b using a left to right, top to bottom ordering. A 32 bit auto-encoding was used for each of the feature maps resulting in a flattened vector of size 64x32 = 2048x1. The traditional flattening obtained by the pre-trained ResNet50 prior to classification, which also generates a 2048x1 vector, is shown in Figure~\ref{resnet}d. As can be seen by their intensity renderings, the spatially preserving flattening shows more discriminable intensity values, intuitively supplying the basis for improvement in subsequent classification. 
 \subsection{Spatially-preserving deep learning network}
\begin{figure}[htb]
\begin{minipage}[b]{1.0\linewidth}
  \centering
  \centerline{\includegraphics[width=8.5cm]{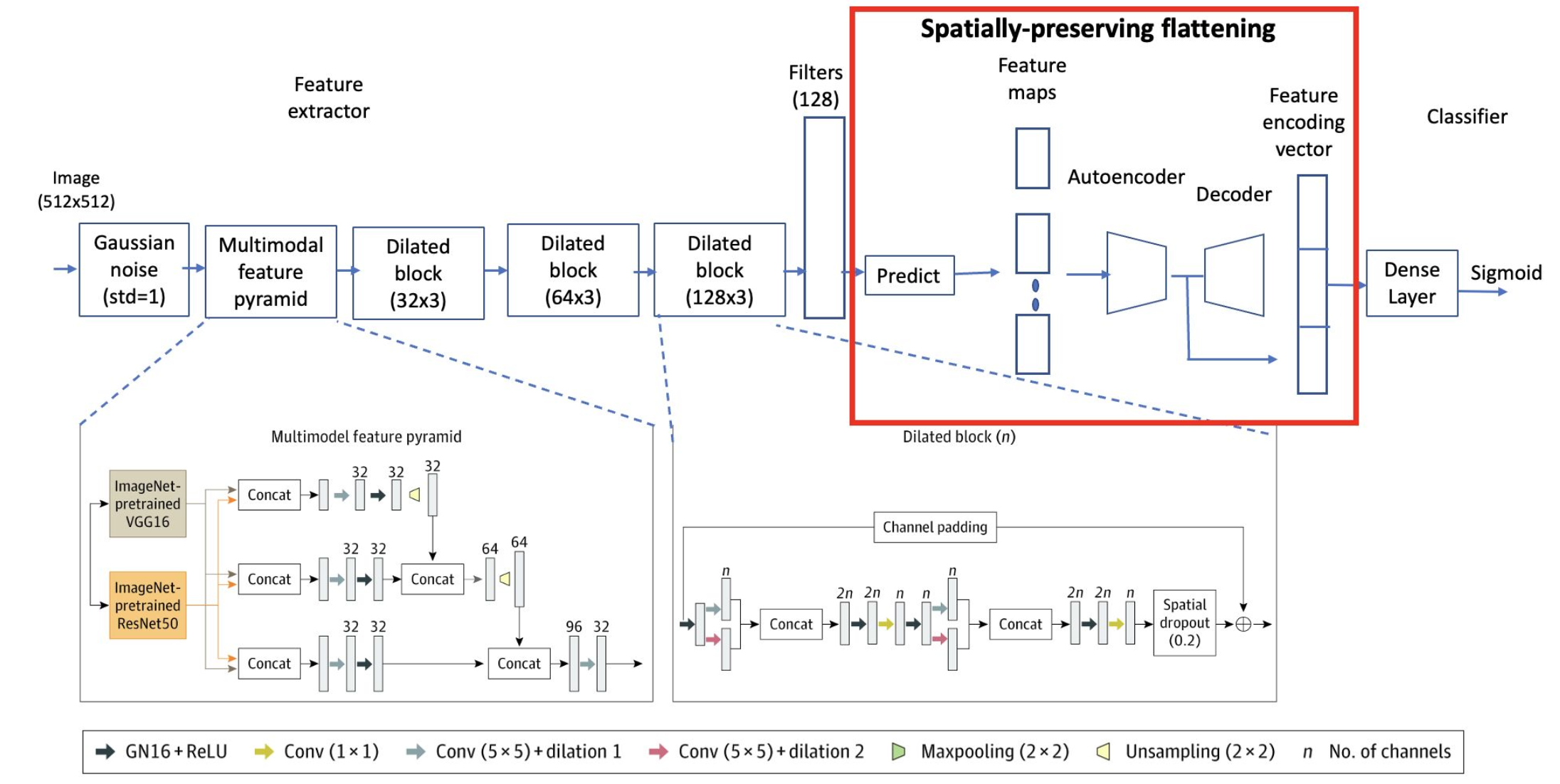}}
\end{minipage}
\caption{Illustration of spatially preserving flattening deep learning network architecture.}
\label{architecture}
\end{figure}

Although all feature maps from the second layer were used in the above example, 
%new deep learning architectures can be formed from any layer and even by mixing across layers using their autoencodings to form the flattening layer. While this could result in fewer layer architectures in general, the focus of this paper is to show the utility of the spatially preserving flattening in improving the classification of location-specific finding in chest X-rays. 
it is preferred to use the last layer of the feature extraction step to benefit from the local and global image characteristics captured in the filters. %Specifically, we chose a configuration in which the spatially preserving flattening is derived from the filters of the last layer of the feature extraction stage. 
Specifically, we introduce a spatially preserving flattening module in between the feature extraction and classification layers to make a new end-to-end deep learning network as shown in Figure~\ref{architecture}. The proposed network leverages the advantages of the feature extractor and classifier portion of a previously developed network\cite{tanveer2020}. Referring to \cite{tanveer2020} for details, we describe these stages only briefly here. %In particular, gaussian noise is added to the incoming chest X-ray images prior to feature extraction to avoid overfitting. 
The feature extractor uses a feature pyramid network to allow multiresolution analysis, and a cascade of dilated blocks with skip connections to improve convergence, while spatial dropout was used to reduce overfitting. Dilated blocks with different feature channels were cascaded with maxpooling to learn more abstract features. 
%Group normalization (16 groups) was used and rectified linear unit served for activation functions. 
Finally the classification was done through a dense layer with sigmoid activation to allow multilabel classification. The parameters and dimensions are detailed in Figure~\ref{architecture}. The last stage of the feature extractor generates 128 filters of size 64x64. In the original architecture described in \cite{tanveer2020}, global square pooling was used to flatten these filter weights into a feature vector of size 128*128 = 16,384.  
The new spatially preserving flattening (SPF) module aims to replace this 16,384 length vector with one generated from auto-encoded feature maps. Specifically, the new flattening module consists of (a) a predictor to predict activation maps using the 128 filters from the last stage of feature extraction, (b) an autoencoder to train on the feature maps for capturing their spatial information, and (c) a flattener that concatenates the encodings of the feature maps. Specifically, we used a 128-bit encoder for each of the 128 incoming 512x512 feature maps produced by the predictor. The resulting flattened vector was assembled by arranging the 128 encoded vectors of all features maps into a 16,384 length vector which was then given as input to the DenseNet classifier.
\begin{table}
    \begin{small}
    \begin{tabular}{|l|l|l|l|l|}
    \hline
       Label & Images &  \multicolumn{3}{c|}{Macro AUC} \\\cline{3-5}
        & & Ours & Network1 & Detectron\\  
        \hline
        Elevated right & 1596 & 0.892 & 0.797 & 0.654\\
        hemidiaphragm & & & &\\
        \hline
        Bilateral & 6696 & 0.823 & 0.793 & 0.71\\
        pleural effusions & & & &\\
        \hline
        Pneumothorax in & 4442 & 0.882 & 0.784 &0.679\\
        the left lower lobe & & & &\\
        \hline
        Right lower lobe & 44770 & 0.812 & 0.766 & 0.71\\
        pleural effusion & & & &\\
        \hline
    \end{tabular}
    \end{small}
    \caption{Illustration of frequently seen location-specific findings and the performance of various learning networks. }
    \label{auc}
\end{table}

The overall network was trained end-to-end using a Nadam optimizer for fast convergence, with the learning rate as 2x10$^{-6}$. Two NVIDIA Tesla V100 graphics processing units with 16 GB memory were used for multi–graphics processing unit training with a batch size of 12 over 30 epochs. All development was done in Python using TensorFlow and Keras libraries. 
\section{Classification of location-specific findings in chest X-rays}
\label{sec:pagestyle}
We now apply the developed deep learning network for location-specific finding classification in chest X-rays. A set of location-specific labels were catalogued for a large collection of chest X-ray images in a previous work\cite{Wu2020} which was leveraged in our work as well. Specifically, this was a  multi-institutional dataset collected from 2 reference sources,
namely, MIMIC-CXR\cite{Johnson2019}, NIH\cite{WangXPengYLuLLuZBagheriM2017}. These X-rays showed a wide range of clinical settings, including intensive care units, urgent care, in-patient care, and emergency departments and represented a wide variety of chest X-ray findings in AP, PA and portable modes of acquisition. All data used was de-identified and covered under the secondary use of health data and informed patient consent was either waived (NIH) or obtained (MIMIC) by the data providers. An earlier work had painstakingly catalogued a full list of 457 findings in chest X-rays using text analytics on the accompanying radiology reports associated with these images\cite{syeda-mahmood2020}. %This algorithm first locates core findings using a clinician-curated chest x-ray lexicon of terms and a vocabulary-driven textual concept extraction algorithm. Natural language analysis of the sentences using dependency parsing and phrasal grouping algorithm then associates detailed characterization modifiers with the relevant core findings in sentences. 
In our work, we selected a subset of 57 of the 457 labels from this label set that were found to refer to either laterality or anatomical location and were most frequently found in radiology reports. A subset of these labels are shown in Table~\ref{auc}. Since the earlier work\cite{syeda-mahmood2020} had already verified these labels, we used the labeled dataset to train, validate, and test our network for fine-grained finding classification. We used a 70-10-20\% split for training, validation and testing. Although the full dataset consisted of 335,189 chest x-rays and their reports, the dataset available for the 57 finding labels was found to contain 107,169 chest X-ray images. 
\begin{table}
    \begin{tabular}{|l|l|l|l|l|l|}
    \hline
        Method & Labels & Train &  Test & Macro AUC & Weighted AUC\\
        \hline
\cite{Carion2020} &57& 75,613& 20,941& 0.512±0.025 & 0.573±0.022 \\
\cite{Wu2020}& 457 & 75,613& 20,941& 0.729±0.001 &  0.716±0.002\\
\cite{Wu2020} &57 & 75,613& 20,941& 0.74±0.003 &0.723 ±0.012\\
Ours &57 &  75,613 & 20,941 &0.81±0.001& 0.79±0.002\\
\hline
    \end{tabular}
    \caption{Illustration of the performance of location and laterality classification models against the selected dataset using labels. The first two columns list the method and the number of labels.}
    \label{overall}
\end{table}
\section{Results}
\label{sec:typestyle}
\begin{figure}[htb]
\begin{minipage}[b]{1.0\linewidth}
  \centering
  \centerline{\includegraphics[width=8.5cm]{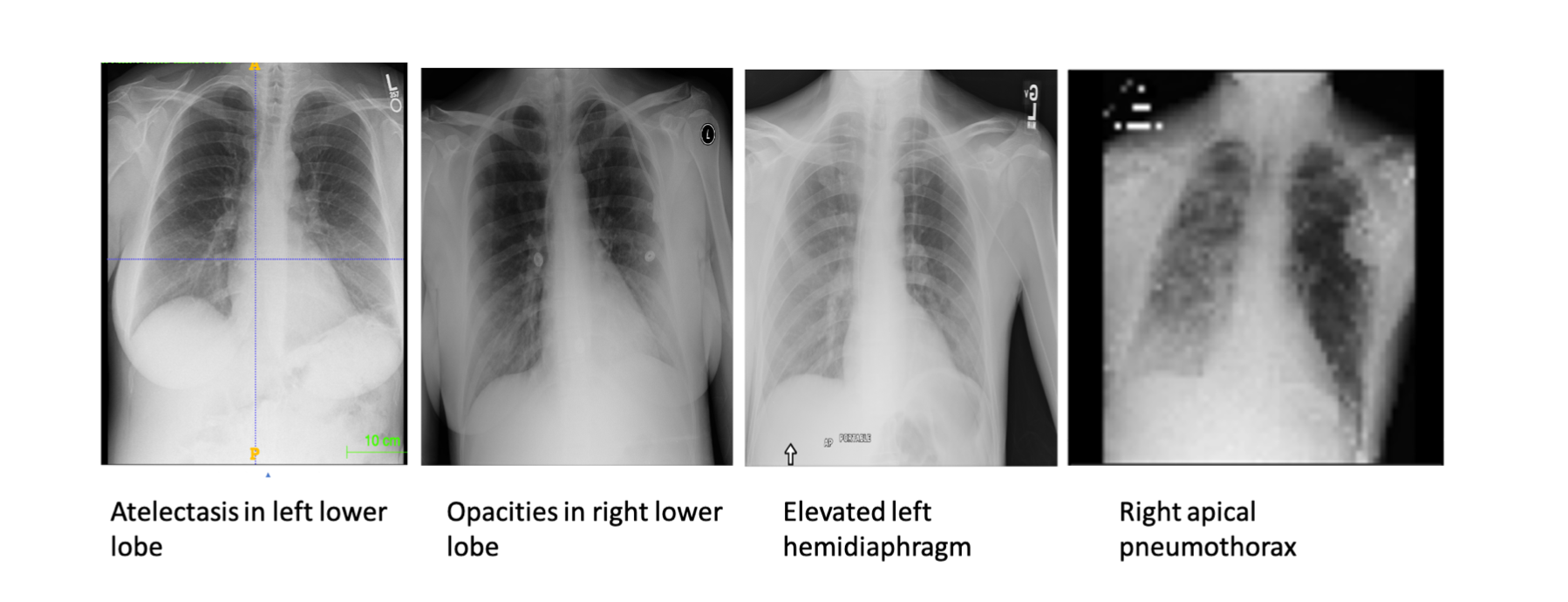}}
\end{minipage}
\caption{Illustration of location specific finding classification using SPF deep learning network.}
\label{res1}
\end{figure}

By using spatially sensitive encodings from feature maps for classification, our network is able to predict the location of specific labels reliably.  Figure~\ref{res1} shows the type of location-specific label prediction on sample images by our network. Since the location-specific details were already included in the label, the AUC measured the combined accuracies of anatomy and anomaly classification. The weighted AUC for the 57 labels was found to be 0.81. 

\noindent{\bf Comparison of performance}

As the literature on location-specific finding classification in chest X-rays is sparse, we compared the performance of our deep learning network against two other state-of-the-art approaches, namely, those using an anatomical atlas to separate anatomy detection from anomaly location, and those that use a whole image approach to directly recognize location-specific labels. Since their networks were trained on different sets of labels, we report our evaluation on the same subset of 57 anatomy-specific findings we chose for our implementation. Specifically, for the atlas approach, we implemented a detection transformer (DETR) with ResNet-50 as base model using the implementation provided from DETR\cite{Carion2020}. This was trained on the automatically extracted bounding box regions obtained by applying the anatomical atlas to the training and testing images as reported in \cite{Wu2020AutomaticBB}. Since no open source implementation of \cite{Agu2021} released this year was provided, the above implementation closely approximates what is possible with SOTA approaches based on anatomical atlases, although the reference\cite{Agu2021} was demonstrated for only 9 core findings. Finally, the network reported in \cite{Wu2020} gave us a comparison to conventional non-spatially preserving encoding-based fine-grained classification. All networks used the same splits for train and test as shown in Table~\ref{overall}, and the same set of 57 labels. As can be seen, the average AUC obtained by our method is at least 15\% higher than the comparable methods. The performance for individual fine-grained labels is shown in Table~\ref{auc}.

\noindent{\bf Ablation studies}

The main ablation study was to see the effect of the use of spatially preserving flattening over conventional flattening. Since the comparison network of \cite{Wu2020} used this form of flattening, the results in Table~\ref{auc} and Table~\ref{overall} constitute the ablation study for the flattening. We also performed an ablation study by changing the size of the auto-encoding for the feature maps and selecting the filters from different layers of the feature extraction stage. The effect of auto-encoding sizes on the average AUC performance is shown in Table~\ref{aucvar}, indicating that the size of the encoding chosen for producing the feature maps was appropriate for the image sizes being handled in our network for chest X-rays. 
\section{Conclusion}
\label{sec:majhead}
\begin{table}
    \centering
    \begin{tabular}{|c|c|}
    \hline
        Encoding size & AUC \\
        \hline
32& 0.67±0.002 \\
64 & 0.716±0.003\\
128 & 0.81±0.012\\
256& 0.80±0.001\\
\hline
    \end{tabular}
    \caption{Illustration of ablation study on the size of the auto-encoding used for capturing the shapes of filters. The train and test splits of the network are the same as in Table~\ref{overall}.}
    \label{aucvar}
\end{table}
In this paper,  we have presented a new spatially preserving deep learning network for location-aware finding classification in chest X-rays. Results are shown on a large multi-hospital chest X-ray dataset indicating a significant improvement in the quality of finding classification over state-of-the-art methods. 
\bibliographystyle{IEEEbib}
\bibliography{strings,refs}
\end{document}